\documentclass[11pt]{article}
\usepackage{longtable}
\usepackage{ae,aecompl,pslatex}
\usepackage[dvips]{graphicx}
\usepackage{enumerate,color}
\usepackage{fullpage}
\usepackage{lineno}
\usepackage[table]{xcolor}
\definecolor{lightgray}{gray}{0.9}
\usepackage{float}

\usepackage{placeins}
\usepackage{psfrag,epsfig,graphicx,color,array,dcolumn,tabularx,delarray,longtable,indentfirst,amsmath,amsfonts,amssymb,amsthm,eucal,bbm}
\theoremstyle{bkaexa} 
\usepackage{caption}
\usepackage{setspace}
\theoremstyle{bkaexa} 

\theoremstyle{bkathm} 

\theoremstyle{bkathm} 
\newtheorem{Thm}{Theorem}
\theoremstyle{bkathm} 

\theoremstyle{bkathm} 
\newtheorem{Lem}{Lemma}
\theoremstyle{definition}

\begin{document}
\setstretch{1.5}
\title{\huge On the asymptotic distributions of some test statistics for two-way contingency tables}
\author{\normalsize Qingyang Zhang\\
\normalsize Department of Mathematical Sciences, University of Arkansas, Fayetteville, AR 72701, USA\\
\normalsize Email: qz008@uark.edu
}
\date{}
\maketitle

\begin{abstract}
Pearson's Chi-square test is a widely used tool for analyzing categorical data, yet its statistical power has remained theoretically underexplored. Due to the difficulties in obtaining its power function in the usual manner, Cochran (1952) suggested the derivation of its Pitman limiting power, which is later implemented by Mitra (1958) and Meng \& Chapman (1966). Nonetheless, this approach is suboptimal for practical power calculations under fixed alternatives. In this work, we solve this long-standing problem by establishing the asymptotic normality of the Chi-square statistic under fixed alternatives and deriving an explicit formula for its variance. For finite samples, we suggest a second-order expansion based on the multivariate delta method to improve the approximations. As a further contribution, we obtain the power functions of two distance covariance tests. We apply our findings to study the statistical power of these tests under different simulation settings.
\end{abstract}


\section{Introduction}
Pearson's Chi-square test is a common statistical tool for analyzing two-way contingency tables, allowing researchers to assess whether there is a significant association between two categorical variables (Pearson, 1900). Pearson's statistic compares the observed frequencies in each cell of the table to the expected frequencies if there were no association between the variables. Since its first appearance in 1900, Pearson's Chi-square statistic has become one of the best-known and most important objects in statistical analyses, widely employed by scientists across various fields. We begin with a brief review of this important statistic and some relevant results. Let $X$ and $Y$ be two categorical variables, with $X\in\{1,~ ...,~ I\}$ and $Y\in\{1,~ ...,~ J\}$. Let $\pi = (\pi_{ij})_{1\leq i\leq I, 1\leq j\leq J}$ be the joint distribution of $(X, Y)$, where $\pi_{ij} = P(X = i, Y = j)$, and $(\pi_{i+})_{1\leq i\leq I}$ and $(\pi_{+j})_{1\leq j\leq J}$ be the marginal probabilities. When a sample of $n$ observations are classified with respect to $X$ and $Y$, the resulting frequencies are often displayed in a two-way contingency table of dimension $(I,~J)$. Let $n_{ij}$ be the observed count in cell $(i, j)$, $n_{i+} = \sum_{j=1}^{J}n_{ij}$ and $n_{+j} = \sum_{i=1}^{I}n_{ij}$ be the marginal counts. The maximum likelihood estimates are $\widehat{\pi}_{ij} = n_{ij}/n$, $\widehat{\pi}_{i+} = n_{i+}/n$, and $\widehat{\pi}_{+j} = n_{+j}/n$. The hypothesis testing for independence between $X$ and $Y$ can be formulated as 
\begin{align*}
H_{0}: \pi_{ij} &= \pi_{i+}\pi_{+j}\text{~for~}1\leq i\leq I,~ 1\leq j\leq J, \\
H_{a}: \pi_{ij} &\neq \pi_{i+}\pi_{+j}\text{~for~some~}(i, j).
\end{align*}
Pearson's Chi-square statistic can be expressed as 
\begin{equation*}
\Delta_{n} = \sum_{i=1}^{I}\sum_{j=1}^{J}\frac{(\widehat{\pi}_{ij} - \widehat{\pi}_{i+}\widehat{\pi}_{+j})^2}{\widehat{\pi}_{i+}\widehat{\pi}_{+j}},
\end{equation*} 
which is the maximum likelihood estimate of the following functional
\begin{equation*}
\Delta = \sum_{i=1}^{I}\sum_{j=1}^{J}\frac{(\pi_{ij} - \pi_{i+}\pi_{+j})^2}{\pi_{i+}\pi_{+j}}.
\end{equation*} 
It is well-known that under the null hypothesis of independence, $n\widehat{\Delta}_{n}$ converges to a Chi-square distribution with $df = (I-1)(J-1)$. As a practical guideline, it is recommended that the expected frequencies, i.e., $n\widehat{\pi}_{i+}\widehat{\pi}_{+j}$, are at least 5. For small expected frequencies, Cochran (1954) introduced a normal approximation to the null distribution of $n\widehat{\Delta}_{n}$, using the exact mean and variance derived by Haldane (1940). Alternative gamma and lognormal approximations were studied by Nass (1959) and Lawal \& Upton (1984). Lewis et al. (1984) derived the third central moment and studied a three-moment location-shifted gamma approximation. Rempala \& Wesolowski (2016) established double asymptotics (Poissonian and Gaussian) of Pearson's Chi-square statistic for one-dimension goodness-of-fit.

Despite a rich literature on improved approximations of $n\widehat{\Delta}_{n}$ under the null hypothesis, its asymptotic behavior under alternative hypotheses remains relatively unexplored. Due to the difficulties in obtaining its asymptotic distribution under fixed alternatives, Cochran (1952) suggested the derivation of its limiting power under Pitman's local alternative, i.e.,
\begin{equation}
\pi_{ij} = \pi_{i+}\pi_{j+} + \frac{c_{ij}}{\sqrt{n}},~\text{where~} \sum_{i=1}^{I}\sum_{j=1}^{J}c_{ij} = 0.
\end{equation}
Later, Mitra (1958) and  Meng \& Chapman (1966) independently showed that under Pitman's local alternative, $n\widehat{\Delta}_{n}$ converges to a non-central Chi-square distribution with $df = (I-1)(J-1)$ and non-centrality parameter 
\begin{equation*}
ncp = \sum_{i=1}^{I}\sum_{j=1}^{J}\frac{c_{ij}^{2}}{\pi_{i+}\pi_{j+}}.
\end{equation*} 
While local alternatives have been employed for practical power analyses, such as in the sample size and power calculations of the log-rank test (Strawderman, 1997), their applicability is limited by the assumption of decreasing dependence strength with increasing sample size. To the best of the author's knowledge, the asymptotic distribution of Pearson's Chi-square statistic for two-way tables under fixed alternatives remains an open problem. To address this long-standing gap, this paper establishes the asymptotic normality of Pearson's Chi-square statistic under fixed alternatives. Specifically, we derive an explicit formula for the asymptotic variance using the delta method. For finite samples, we propose an improved approximation based on the second-order multivariate delta method. Our simulations demonstrate the good performance of the second-order expansions under various models and sample sizes.

As a further contribution, we derive the asymptotic distributions of two distance covariance statistics (Sz\'{e}kely et al., 2007) under fixed alternatives, along with the second-order expansions for improved finite-sample approximations. The two statistics are both based on the following (squared) distance covariance functional, which shares conceptual similarities with $\Delta$ but omits $\pi_{i+}\pi_{+j}$'s on the denominators
\begin{equation*}
D = \sum_{i=1}^{I}\sum_{j=1}^{J}(\pi_{ij} - \pi_{i+}\pi_{+j})^2.
\end{equation*} 

The first statistic of interest is the maximum likelihood estimate of $D$, denoted by $\widehat{D}_{n}$, and defined as follows
\begin{equation*}
\widehat{D}_{n} = \sum_{i=1}^{I}\sum_{j=1}^{J}(\widehat{\pi}_{ij} - \widehat{\pi}_{i+}\widehat{\pi}_{+j})^2.
\end{equation*} 
Zhang (2019) demonstrated that the permutation test based on $\widehat{D}_{n}$ exhibits significantly greater power than Pearson's Chi-square test for insufficient sample sizes. The second statistic that we investigated is an unbiased estimate of $D$, as proposed by Berrett \& Samworth (2021). Their estimator, $\widetilde{D}_{n}$, is derived using a fourth-order U-statistic. Similar to $\widehat{D}_{n}$, the permutation test based on $\widetilde{D}_{n}$ generally outperforms Pearson's Chi-square test.
\begin{align*}
\widetilde{D}_{n} & = \frac{n}{n-3} \sum_{i = 1}^{I} \sum_{j=1}^{J} (\widehat{\pi}_{ij} - \widehat{\pi}_{i+}\widehat{\pi}_{+j})^{2} - \frac{4n}{(n-2)(n-3)}\sum_{i = 1}^{I} \sum_{j=1}^{J}\widehat{\pi}_{ij}\widehat{\pi}_{i+}\widehat{\pi}_{+j}  \\
& + \frac{n}{(n-1)(n-3)}\left( \sum_{i = 1}^{I}\widehat{\pi}_{i+}^{2} + \sum_{j=1}^{J}\widehat{\pi}_{+j}^{2} \right) + \frac{n(3n-2)}{(n-1)(n-2)(n-3)}\left(\sum_{i = 1}^{I}\widehat{\pi}_{i+}^{2}\right)\left(\sum_{j=1}^{J}\widehat{\pi}_{+j}^{2}\right)  \\
& - \frac{n}{(n-1)(n-3)},
\end{align*}
It is noteworthy that $\widetilde{D}_{n}$ is a U-statistic, thus its asymptotic normality under fixed alternatives can be easily established. However, as a fourth-order U-statistic, the derivation of its asymptotic variance can be intricate. Futhermore, the U-statistic theory generally lacks higher-order asymptotic expansions for finite samples. Given these considerations, we employ the multivariate delta method for the asymptotic analysis of $\widehat{D}_{n}$ and $\widetilde{D}_{n}$, as it offers close-form asymptotic variances and higher-order expansions. 

The remainder of this paper is structured as follows: Section 2 presents the asymptotic distributions of $\Delta_{n}$, $\widehat{D}_{n}$ and $\widetilde{D}_{n}$, under fixed alternatives, followed by improved approximations derived using the second-order multivariate delta method. Section 3 evaluates the convergence performance of these asymptotic distributions through simulations under various models. Section 4 concludes the paper with a discussion and some future directions.   

\section{The asymptotic distributions of $\Delta_{n}$, $\widehat{D}_{n}$ and $\widetilde{D}_{n}$}
\subsection{Under the null hypothesis}
Under the null hypothesis of independence, with fixed number of categories, $I$ and $J$, and the joint probabilities $\pi_{ij}$'s, it is well-known that $n\Delta_{n}$ asymptotically follows a Chi-square distribution with $df = (I-1)(J-1)$. Zhang (2024) derived the asymptotic null distribution of the unbiased estimate of squared distance covariance, given by
\begin{equation}\label{unbias}
n\widetilde{D}_{n} \xrightarrow{d} \sum_{i=1}^{I-1}\sum_{j=1}^{J-1} \lambda_{i}\gamma_{j}(Z^{2}_{ij}-1),
\end{equation}
where $Z_{ij}$'s are independent standard normal random variables, and $\{\lambda_{1},~...,~\lambda_{I-1}\}$ and $\{\gamma_{1},~...,~\gamma_{J-1}\}$ are the non-zero eigenvalues of two matrices (of rank $I-1$ and $J-1$, respectively) that depend solely on the marginal probabilities. These matrices are explicitly defined in Zhang (2024), Section 4. 

Zhang (2024) also gave the close-form expressions of  $\{\lambda_{1},~...,~\lambda_{I-1}\}$ and $\{\gamma_{1},~...,~\gamma_{J-1}\}$ for small tables. For instance, in the special case of $I=2$, we have $\lambda_{1} = -2\pi_{1+}\pi_{2+}$. For $I = 3$, we have
\begin{align*}
\lambda_{1} & = -(\pi_{1+}\pi_{2+}+\pi_{1+}\pi_{3+} + \pi_{2+}\pi_{3+}) - \sqrt{\pi_{1+}^{2}\pi_{2+}^{2}+\pi_{1+}^{2}\pi_{3+}^{2} + \pi_{2+}^{2}\pi_{3+}^{2}-\pi_{1+}\pi_{2+}\pi_{3+}}, \\
\lambda_{2} & = -(\pi_{1+}\pi_{2+}+\pi_{1+}\pi_{3+} + \pi_{2+}\pi_{3+}) + \sqrt{\pi_{1+}^{2}\pi_{2+}^{2}+\pi_{1+}^{2}\pi_{3+}^{2} + \pi_{2+}^{2}\pi_{3+}^{2}-\pi_{1+}\pi_{2+}\pi_{3+}}.
\end{align*}

For the maximum likelihood estimate $\widehat{D}_{n}$, Edelmann \& Goeman (2022) and Castro-Prado et al. (2024) derived its asymptotic null distribution, which is a weighted sum of Chi-square variables. Motivated by Edelmann \& Goeman (2022), Castro-Prado et al. (2024) and Zhang (2024), we investigated the limiting distribution of $n(\widehat{D}_{n} - \widetilde{D}_{n})$. Lemma \ref{lem1} below demonstrates that $n(\widehat{D}_{n}-\widetilde{D}_{n})$ converges almost surely to a constant (see Appendix A.1 for the proof). Under independence, this constant reduces to a product of two entropy measures.

\begin{Lem} \label{lem1}
For categorical variables $X$ and $Y$ with joint probabilities $\pi_{ij}$'s
\begin{equation*}
n(\widehat{D}_{n} - \widetilde{D}_{n}) \xrightarrow{a.s.}  1 - \sum_{i=1}^{I}\pi^{2}_{i+} - \sum_{j=1}^{J}\pi^{2}_{+j} - 3\sum_{i=1}^{I}\sum_{j=1}^{J}\pi^{2}_{i+}\pi^{2}_{+j} + 4\sum_{i=1}^{I}\sum_{j=1}^{J}\pi_{ij}\pi_{i+}\pi_{+j} - 3D ,
\end{equation*}
as $n\rightarrow\infty$. If $X$ and $Y$ are independent, we have
\begin{equation*}
n(\widehat{D}_{n} - \widetilde{D}_{n}) \xrightarrow{a.s.} \left(1-\sum_{i=1}^{I}\pi^{2}_{i+}\right)\left(1-\sum_{j=1}^{J}\pi^{2}_{+j}\right).
\end{equation*}
\end{Lem}

By Lemma \ref{lem1}, Equation \ref{unbias} and Slutsky's theorem, it follows directly that
\begin{equation*}
n\widehat{D}_{n} \xrightarrow{d} \sum_{i=1}^{I-1}\sum_{j=1}^{J-1} \lambda_{i}\gamma_{j}(Z^{2}_{ij}-1) + \left(1-\sum_{i=1}^{I}\pi^{2}_{i+}\right)\left(1-\sum_{j=1}^{J}\pi^{2}_{+j}\right),
\end{equation*}
which can be an alternative expression of the results by Edelmann \& Goeman (2022) and Castro-Prado et al. (2024). Lemma \ref{lem1} is also essential for proving our main theorems, presented in Section 2.2.

\subsection{Under alternative hypotheses}
Under Pitman's local alternatives, Mitra (1958) and Meng \& Chapman (1966) showed that $n\widehat{\Delta}_{n}$ converges in distribution to a non-central Chi-square distribution. They subsequently applied this result to power calculations. Nonetheless, this approach can be problematic for calculating power under fixed alternatives, which are more common in practice. Furthermore, our simulation study demonstrates that the non-central Chi-square distribution often fails to provide accurate approximations, even under local alternatives.

Here we present the asymptotic distributions of the three statistics under fixed alternatives. To begin, we introduce some notations. Let $\pi = (\pi_{ij})$ be the matrix of joint probabilities. For an $I$-by-$J$ matrix $A$, let $\mathrm{vec}^{*}(A)$ denote the leave-one-out vectorization of $A$, obtained by stacking all the columns and removing the last element $A_{IJ}$. We denote by $\Sigma^{*}$ the variance-covariance matrix of $\mathrm{vec}^{*}(\sqrt{n}\widehat{\pi})$, where $\mathrm{Var}(\sqrt{n}\widehat{\pi}_{ij}) = \pi_{ij}(1-\pi_{ij})$ and $\mathrm{Cov}(\sqrt{n}\widehat{\pi}_{ij},~\sqrt{n}\widehat{\pi}_{km}) =  - \pi_{ij} \pi_{km}$ for $i\neq k$ or $j\neq m$. For instance, for $I=3$ and $J=4$, $\Sigma^{*}$ is a 11-by-11 matrix. In addition, let $D' = (D'_{ij})$ and $\Delta ' = (\Delta '_{ij})$ be the matrices of first-order partial derivatives, where $D'_{ij} = \partial D / \partial \pi_{ij}$ and $\Delta '_{ij} = \partial \Delta / \partial \pi_{ij}$.

The following main theorems establish the asymptotic normality of $\widehat{D}_{n}$, $\widetilde{D}_{n}$ and $\widehat{\Delta}_{n}$, along with close-form asymptotic variances. Notably, our Theorems \ref{thm1} and \ref{thm2} indicate that the asymptotic variances of all three statistics depend on not only the dependence strength (quantified by $c_{ij}$'s), but also on the marginal probabilities. The proofs of both theorems employ the multivariate delta method (see Appendix A.2 and A.3 for details). To the best of our knowledge, this is the first derivation of asymptotic normality and closed-form asymptotic variances for these quantities under fixed alternatives.
\begin{Thm}\label{thm1}
Under fixed alternatives, $\pi_{ij} = \pi_{i+}\pi_{+j} + c_{ij}$, where $c_{ij}$'s are constants that satisfy $\sum_{i=1}^{I}\sum_{j=1}^{J}c_{ij} = 0$ and $\max_{i, j} |c_{ij}| > 0$, we have
\begin{equation*}
\sqrt{n}(\widehat{\Delta}_{n} - \Delta)  \xrightarrow{d} N\left\{ 0, ~ [ \mathrm{vec}^{*}(\Delta') ]^{\intercal} \Sigma^{*} \mathrm{vec}^{*}(\Delta') \right\}, 
\end{equation*}
where  
$$
\Delta'_{ij} = 
\begin{cases}
\sum_{k=1}^{J}\frac{2c^{2}_{Ik}}{\pi^{2}_{I+}\pi_{+k}} + \sum_{m=1}^{I}\frac{2c^{2}_{mJ}}{\pi_{m+}\pi^{2}_{+J}} - \sum_{k=1}^{J}\frac{2c^{2}_{ik}}{\pi^{2}_{i+}\pi_{+k}} -  \sum_{m=1}^{I}\frac{2c^{2}_{mj}}{\pi_{m+}\pi^{2}_{+j}} + \frac{2c_{ij}}{\pi_{i+}\pi_{+j}} - \frac{2c_{IJ}}{\pi_{I+}\pi_{+J}}, & \text{  for } i \neq I, j \neq J \\
\sum_{m=1}^{I}\frac{2c^{2}_{mJ}}{\pi_{m+}\pi^{2}_{+J}} -  \sum_{m=1}^{I}\frac{2c^{2}_{mj}}{\pi_{m+}\pi^{2}_{+j}} + \frac{2c_{Ij}}{\pi_{I+}\pi_{+j}} - \frac{2c_{IJ}}{\pi_{I+}\pi_{+J}},  & \text{  for } i = I, j \neq J  \\
\sum_{k=1}^{J}\frac{2c^{2}_{Ik}}{\pi^{2}_{I+}\pi_{+k}} -  \sum_{k=1}^{J}\frac{2c^{2}_{ik}}{\pi^{2}_{i+}\pi_{+k}} + \frac{2c_{iJ}}{\pi_{i+}\pi_{+J}} - \frac{2c_{IJ}}{\pi_{I+}\pi_{+J}}, & \text{  for } i \neq I, j = J 
  \end{cases}  
  $$
\end{Thm}

\begin{Thm}\label{thm2}
Under the conditions of Theorem \ref{thm1}, we have
\begin{align*}
\sqrt{n}(\widehat{D}_{n} - D) & \xrightarrow{d} N\left\{ 0, ~ [ \mathrm{vec}^{*}(D') ]^{\intercal} \Sigma^{*} \mathrm{vec}^{*}(D') \right\}, \\
\sqrt{n}(\widetilde{D}_{n} - D) & \xrightarrow{d} N\left\{ 0, ~  [ \mathrm{vec}^{*}(D') ]^{\intercal} \Sigma^{*} \mathrm{vec}^{*}(D') \right\},
\end{align*}
as $n\rightarrow\infty$, where 
$$
D'_{ij} = 
\begin{cases}
2\sum_{m=1}^{I}\pi_{m+}c_{mJ} + 2\sum_{k=1}^{I}\pi_{+k}c_{Ik} - 2\sum_{m=1}^{I}\pi_{m+}c_{mj} - 2\sum_{k=1}^{I}\pi_{+k}c_{ik} + 2c_{ij} - 2c_{IJ}, & \text{  for } i \neq I, j \neq J \\
2\sum_{m=1}^{I}\pi_{m+}c_{mJ} - 2\sum_{m=1}^{I}\pi_{m+}c_{mj} + 2c_{Ij} - 2c_{IJ}, & \text{  for } i = I, j \neq J  \\
2\sum_{k=1}^{I}\pi_{+k}c_{Ik} - 2\sum_{k=1}^{I}\pi_{+k}c_{ik} + 2c_{iJ} - 2c_{IJ}, & \text{  for } i \neq I, j = J 
 \end{cases}  
  $$
\end{Thm}

In the above theorems, it is straightforward to verify that $\Sigma^{*}$ is positive definite, therefore the asymptotic variances are strictly positive as long as $\max_{i, j} |c_{ij}| > 0$. Under the null hypothesis of independence, all $c_{ij}$'s are zero, leading to zero first-order partial derivatives, i.e., $D'_{ij}=\Delta'_{ij}=0$ for all $(i,~j)$'s. This results in degenerate statistics of weighted Chi-square forms (see Equation \ref{unbias}). 

As Theorem \ref{thm2} demonstrates, the two distance covariance statistics are asymptotically equivalent under fixed alternatives. This may initially seem contradictory to Lemma \ref{lem1}, but it is not. By Lemma \ref{lem1}, the difference between the two statistics, i.e., $\widehat{D}_{n} - \widetilde{D}_{n}$ almost surely converges to a constant at a rate of $1/n$. Consequently, $\sqrt{n}(\widehat{D}_{n} - \widetilde{D}_{n})$ almost surely converges to 0, leading to the asymptotic equivalence between $\sqrt{n}(\widehat{D}_{n}-D)$ and $\sqrt{n}(\widetilde{D}_{n}-D)$. 

It is also important to note that our asymptotic results may not always accurately approximate the power functions for finite samples, particularly with certain distributions, as illustrated in our simulation study (Section 3). To address this, we provide the second-order asymptotic expansions below, which can significantly improve the approximations for finite samples. The proof of Theorem \ref{thm3} is given in Appendix A.4 (the proof of Theorem \ref{thm4} is omitted as it is similar and follows directly from Lemma \ref{lem1}).

\begin{Thm}\label{thm3}
Under fixed alternatives and finite samples, we have
\begin{equation*}
\sqrt{n}(\widehat{\Delta}_{n} - \Delta)  \approx \sigma_{\Delta} Z_{0} + \frac{1}{2\sqrt{n}} \sum_{g = 1}^{IJ-1} \beta_{g} Z_{g}^{2}, 
\end{equation*}
where $Z_{0},~Z_{1},~...,~Z_{IJ-1}$ are $i.i.d.$ standard normal random variables, and $\sigma^{2}_{\Delta} = [ \mathrm{vec}^{*}(\Delta') ]^{\intercal} \Sigma^{*} \mathrm{vec}^{*}(\Delta')$. The weights $\{\beta_{1},~...,~\beta_{G}\}$ are the eigenvalues of $(\Sigma^{*})^{1/2}H^{*}(\Delta)(\Sigma^{*})^{1/2}$, where $H^{*}(\Delta)$ is the Hessian matrix of $\Delta$ (excluding $\pi_{IJ}$).
\end{Thm}

\begin{Thm}\label{thm4}
Under fixed alternatives and finite samples, we have
\begin{align*}
\sqrt{n}(\widehat{D}_{n} - D) &  \approx \sigma_{D} Z_{0}  + \frac{1}{2\sqrt{n}} \sum_{g = 1}^{IJ-1} \beta_{g} Z_{g}^{2}, \\
\sqrt{n}(\widetilde{D}_{n} - D)  & \approx \sigma_{D} Z_{0} + \frac{1}{2\sqrt{n}} \sum_{g = 1}^{IJ-1} \beta_{g} Z_{g}^{2} - \frac{1}{\sqrt{n}}\left(1 - \sum_{i=1}^{I}\pi^{2}_{i+} - \sum_{j=1}^{J}\pi^{2}_{+j} - 3\sum_{i=1}^{I}\sum_{j=1}^{J}\pi^{2}_{i+}\pi^{2}_{+j} + 4\sum_{i=1}^{I}\sum_{j=1}^{J}\pi_{ij}\pi_{i+}\pi_{+j} - 3D \right),
\end{align*}
where $Z_{0},~Z_{1},~...,~Z_{IJ-1}$ are $i.i.d.$ standard normal random variables, and $\sigma^{2}_{D} = [ \mathrm{vec}^{*}(D') ]^{\intercal} \Sigma^{*} \mathrm{vec}^{*}(D')$. The weights $\{\beta_{1},~...,~\beta_{G}\}$ are the eigenvalues of $(\Sigma^{*})^{1/2}H^{*}(D)(\Sigma^{*})^{1/2}$, where $H^{*}(D)$ is the Hessian matrix of $D$ (excluding $\pi_{IJ}$).
\end{Thm}

Theorems \ref{thm3} and \ref{thm4} establish that for all three statistics, we can enhance accuracy by employing their second-order approximations. These approximations can be represented as a weighted sum of a standard normal variable and $i.i.d.$ Chi-square random variables with $1$ degree of freedom. While such distributions generally lack closed-form density functions or cumulative distribution functions, numerous software packages, e.g, the \textit{psum.chisq} function from R package \textit{mgcv}, offer efficient approximations. Our simulation studies (Section 3) confirmed the efficacy of the second-order formulas, demonstrating significant improvements over the asymptotic distributions. Deriving the Hessian matrices, $H^{*}(\Delta)$ and $H^{*}(D)$, is still possible but complicated. As such, we advocate for a numerical method. For example, in our simulation study, we utilized the \textit{hessian} function in R package \textit{numDeriv}, which is both accurate and efficient.

\section{Simulation studies}
In this section, we assess the accuracy of the approximations derived in Section 2, including the asymptotic distributions in Theorems \ref{thm1} and \ref{thm2}, as well as the second-order expansions presented in Theorems \ref{thm3} and \ref{thm4}. To facilitate our analysis, we employed two simulation models adapted from Berrett \& Samworth (2021).
\begin{itemize}
\item Setting 1: $I = J = 6$, and $\pi_{ij} = \pi^{0}_{ij} + \epsilon (-1)^{(i+j)}$, where $\pi^{0}_{ij} = 1/IJ$. 
\item Setting 2: $I = J = 4$, and
$$
\pi_{ij} = 
\begin{cases}
\pi^{0}_{ij} + \epsilon,  & \text{  for } (i, j) = (1, 1),  (2, 2) \\
\pi^{0}_{ij} - \epsilon,  & \text{  for } (i, j) = (1, 2), (2, 1)  \\
\pi^{0}_{ij},  & \text{  otherwise}. 
\end{cases}  ,
$$where
\begin{equation*}
\pi^{0}_{ij} = \frac{2^{-(i+j)}}{\left( 1-2^{-I} \right)\left( 1-2^{-J} \right)}
\end{equation*}
\end{itemize}
It is straightforward to verify that for both settings, $X$ and $Y$ are independent if and only if $\epsilon=0$. We have a fixed alternative if $\epsilon$ is a non-zero constant, and Pitman's local alternative if $\epsilon = 1/\sqrt{n}$ (assuming $c_{ij} = 1$ without loss of generality). Setting 1 represents a scenario with uniformly distributed marginal probabilities and widespread dependence across all cells. In contrast, Setting 2 exhibits non-uniform marginal probabilities and dependence confined to a specific subset of cells. Figure 1 provides visual examples for each setting, wherein $\epsilon = 1/40$ for Setting 1 and $\epsilon = 1/10$ for Setting 2.  

\begin{center}
[Figure 1 about here]
\end{center}

First, we investigated the distributions of the three statistics under Pitman's local alternatives, uncovering some instances of slow convergence. Focusing on Pearson's Chi-square statistic (with analogous results for the two distance covariance statistics), Figure 2 illustrates the distribution of $n\Delta_{n}$ based on 100,000 simulations. Interestingly, it is found that $n\Delta_{n}$ has different convergence rates across the two settings. In Setting 2, $n\Delta_{n}$ converges to its asymptotic distribution at a sample size of 200 (with an average cell count of 12.5). However, in Setting 1, the finite-sample distribution remains significantly distinct from the asymptotic distribution even with a large sample size of $n = 5,000$ (average cell count of 138.9). Our simulation findings emphasize the need for caution when applying asymptotic results under local alternatives (such as those from Mitra, 1958) in practical sample size and power calculations, as they may yield inaccurate approximations in certain scenarios.

\begin{center}
[Figure 2 about here]
\end{center}

Next, we evaluated the finite-sample performance of the asymptotic results under fixed alternatives (as presented in Theorems \ref{thm1}, \ref{thm2}, \ref{thm3} and \ref{thm4}). Figures 3-5 summarize the distributions of $\Delta_{n}$, $\widehat{D}_{n}$ and $\widetilde{D}_{n}$ based on 10,000 simulations, where the dependence strength parameters are chosen as $\epsilon = 1/40$ for Setting 1 and $\epsilon = 1/10$ for Setting 2, and sample sizes are set to be $n=200,~1000,~5000$. All three statistics exhibit substantially different convergence rates in the two settings, with generally faster convergence observed in Setting 2 (similar to the behavior under local alternatives). For instance, in Setting 2, the unbiased estimate $\widetilde{D}_{n}$ converges to its asymptotic distribution (red dashed) even with a relatively small sample size of $n=200$. However, in Setting 1, its convergence is significantly slower, with a sample size of $n=5,000$ still insufficient. Conversely, it is found that the use of our second-order expansion formulas can significantly improve the approximations. For instance, across all three statistics and both settings, $n=200$ is generally adequate for employing a second-order formula to achieve a relatively accurate approximation.

\begin{center}
[Figure 3 about here]
\end{center}

\begin{center}
[Figure 4 about here]
\end{center}

\begin{center}
[Figure 5 about here]
\end{center}

Finally, we conducted a simulation study to assess the performance of the second-order approximation in power calculations. The sample sizes were chosen to be $\{100,~150,~200,~250\}$ for both settings. The dependence strength parameter $\epsilon$ was selected to be $\{1/100,~1/80\}$ for Setting 1, and $\{1/20,~1/15\}$ for Setting 2. For the theoretical power, all eigenvalues were computed using R function \textit{eigen}, inverse matrices using \textit{solve}, and the Hessian matrices using \textit{hessian} from the \textit{numDeriv} package. The second-order expansions were implemented using the \textit{psum.chisq} function from the \textit{mgcv} package. Tables 1 and 2 compare the theoretical power with the simulation-based power (over 10,000 simulations) for the three tests. We observe that our theoretically calculated power closely approximates the simulated power, with a difference of less than 3\% for most settings. For example, in Setting 1 with $\epsilon = 1/100$ and $n=100$, the unbiased distance covariance test exhibits a simulation-based power of 0.468, compared to the theoretical power of 0.485. Additionally, we note that the two distance covariance tests are generally more powerful than Pearson's Chi-square test, particularly for relatively small sample sizes. For instance, in Setting 2 with $\epsilon = 1/20$ and $n=100$, Pearson's Chi-square test has a statistical power of 0.450 (simulation-based), compared to 0.726 and 0.721 for the two distance covariance tests.

\begin{center}
[Table 1 about here]
\end{center}

\begin{center}
[Table 2 about here]
\end{center}

\section{Discussion and conclusions}
The power of Pearson's Chi-square test for two-way contingency tables remains a long-standing challenge in statistical analysis. In this paper, we address this open problem by establishing the asymptotic distribution of Pearson's Chi-square statistic under fixed alternatives, along with an explicit formula for asymptotic variance. To enhance the approximation for finite samples, we further introduce a second-order expansion. Analogous results are also derived for two distance covariance tests. Our simulation studies demonstrate the robustness and efficacy of our derived formulas across various distributions.

We explore some potential extensions and limitations of our work. First, while our primary focus has been on two-way tables, our findings can be readily extended to $K$-way tables. For instance, when $K=3$, Pearson's Chi-square statistic can be expressed as
\begin{equation*}
\widehat{\Delta}_{n} = \sum_{i=1}^{I}\sum_{j=1}^{J}\sum_{k=1}^{K}\frac{(\widehat{\pi}_{ijk} - \widehat{\pi}_{i++}\widehat{\pi}_{+j+}\widehat{\pi}_{++k})^2}{\widehat{\pi}_{i++}\widehat{\pi}_{+j+}\widehat{\pi}_{++k}}.
\end{equation*}
Under mutual independence, it is well-established that $n\widehat{\Delta}_{n}$ converges to a Chi-square distribution with $df = IJK-I-J-K+2$. Under fixed alternatives, it can be similarly shown that $\sqrt{n}(\Delta_{n}-\Delta)$ is asymptotically normal, though the derivation of the asymptotic variance may be intricate due to the complex partial derivatives for three-way tables. Similar results can potentially be obtained for distance covariance tests assessing mutual independence in $K$-way tables. A significant hurdle lies in deriving the U-statistics for squared distance covariance. For instance, in the case of $K=3$, the regular U-statistic for squared distance covariance is of order 6, and it becomes of order 8 for $K=4$. Working with the maximum likelihood estimate of squared distance covariance when $K\geq 3$ could be more tractable, which we plan to explore in future research.

Second, similar to the power formula derived by Mitra (1958), our method necessitates a complete specification of the joint distributions of the two categorical variables, i.e., $\pi_{ij}$, $1\leq i\leq I$, $1\leq j\leq J$, which can be difficult in real-world applications. While we can employ educated guesses or leverage historical data for rough estimates, Guenther (1977) suggests selecting a set of potential joint distributions and calculating the lower bound of statistical power. Nevertheless, the accuracy of our results hinges on the reasonable specifications for the underlying joint distributions.

Throughout this paper, we have assumed fixed dimensions $I$ and $J$. However, when these dimensions diverge, deriving the asymptotic distributions of Pearson's statistic or distance covariance statistics becomes theoretically challenging, as the multivariate delta distribution does not directly apply. A novel theoretical framework might be necessary to address these challenges.

\section*{Acknowledgement}
\noindent
The work was supported by an NSF DBI Biology Integration Institute (BII) grant (award no. 2119968; PI-Ceballos).

\section*{Appendix}
\section*{A.1. Proof of Lemma \ref{lem1}}
\begin{proof}
First, we have
\begin{align*}
n(\widehat{D}_{n} - \widetilde{D}_{n}) = & -\frac{3n}{n-3} \sum_{i = 1}^{I} \sum_{j=1}^{J} (\widehat{\pi}_{ij} - \widehat{\pi}_{i+}\widehat{\pi}_{+j})^{2} + \frac{4n^2}{(n-2)(n-3)}\sum_{i = 1}^{I} \sum_{j=1}^{J}\widehat{\pi}_{ij}\widehat{\pi}_{i+}\widehat{\pi}_{+j}  \\
& - \frac{n^2}{(n-1)(n-3)}\left( \sum_{i = 1}^{I}\widehat{\pi}_{i+}^{2} + \sum_{j=1}^{J}\widehat{\pi}_{+j}^{2} \right) - \frac{n^{2}(3n-2)}{(n-1)(n-2)(n-3)}\left(\sum_{i = 1}^{I}\widehat{\pi}_{i+}^{2}\right)\left(\sum_{j=1}^{J}\widehat{\pi}_{+j}^{2}\right)  \\
& + \frac{n^2}{(n-1)(n-3)}.
\end{align*}
By the strong law of large numbers, we have $\widehat{\pi}_{ij} \xrightarrow{a.s.} \pi_{ij}$ for any $i$ and $j$. By continuous mapping theorem, we have
\begin{equation*}
n(\widehat{D}_{n} - \widetilde{D}_{n}) \xrightarrow{a.s.}  1 - \sum_{i=1}^{I}\pi^{2}_{i+} - \sum_{j=1}^{J}\pi^{2}_{+j} - 3\sum_{i=1}^{I}\sum_{j=1}^{J}\pi^{2}_{i+}\pi^{2}_{+j} + 4\sum_{i=1}^{I}\sum_{j=1}^{J}\pi_{ij}\pi_{i+}\pi_{+j} - 3D ,
\end{equation*}
as $n\rightarrow\infty$. If $X$ and $Y$ are independent, we have $D=0$ and $\pi_{ij} = \pi_{i+}\pi_{+j}$, the above equation can be simplified to
\begin{equation*}
n(\widehat{D}_{n} - \widetilde{D}_{n}) \xrightarrow{a.s.} \left(1-\sum_{i=1}^{I}\pi^{2}_{i+}\right)\left(1-\sum_{j=1}^{J}\pi^{2}_{+j}\right).
\end{equation*}
\end{proof}

\section*{A.2. Proof of Theorem \ref{thm1}}
\begin{proof}
Our proof is based on multivariate delta method. First, by multivariate central limit theorem, we have 
\begin{equation} \label{mclt}
\sqrt{n}\left[ \mathrm{vec}^{*}(\widehat{\pi}) - \mathrm{vec}^{*}(\pi) \right] \xrightarrow{d} N(\mathbf{0},~ \Sigma^{*})
\end{equation}
It is important to note that $\mathrm{vec}^{*}(\pi)$ is the leave-one-out vectorization, therefore $\Sigma^{*}$ is positive definite, which is needed to show that the asymptotic variance is strictly positive under fixed alternatives $\pi_{ij} = \pi_{i+}\pi_{+j} + c_{ij}$. 

Next, we derive the partial derivative $\Delta '_{ij}$. For $i\neq I$, $j\neq J$, because $\pi_{IJ} = 1 - \sum_{(i, j)\neq (I, J)}\pi_{ij}$, $\pi_{I+} = 1-\sum_{i=1}^{I-1}\pi_{i+}$ and $\pi_{+J} = 1-\sum_{j=1}^{J-1}\pi_{+j}$, $\Delta$ can be divided into the following parts that depend on $\pi_{ij}$
\begin{align*}
\Delta_{1} & =  \sum_{k\neq j,~J} \frac{(\pi_{ik} - \pi_{i+}\pi_{+k})^2}{\pi_{i+}\pi_{+k}}, \\
\Delta_{2} & = \sum_{m\neq i,~I} \frac{(\pi_{mj} - \pi_{m+}\pi_{+j})^2}{\pi_{m+}\pi_{+j}}, \\
\Delta_{3} & = \frac{(\pi_{ij} - \pi_{i+}\pi_{+j})^2}{\pi_{i+}\pi_{+j}}, \\
\Delta_{4} & = \frac{(\pi_{iJ} - \pi_{i+}\pi_{+J})^2}{\pi_{i+}\pi_{+J}}, \\
\Delta_{5} & = \frac{(\pi_{Ij} - \pi_{I+}\pi_{+j})^2}{\pi_{I+}\pi_{+j}}, \\
\Delta_{6} & =  \sum_{k\neq j,~J} \frac{(\pi_{Ik} - \pi_{I+}\pi_{+k})^2}{\pi_{I+}\pi_{+k}}, \\
\Delta_{7} & = \sum_{m\neq i,~I} \frac{(\pi_{mJ} - \pi_{m+}\pi_{+J})^2}{\pi_{m+}\pi_{+J}}, \\
\Delta_{8} & = \frac{(\pi_{IJ} - \pi_{I+}\pi_{+J})^2}{\pi_{I+}\pi_{+J}}.
\end{align*}
The partial derivative of each part with respect to $\pi_{ij}$ can be obtained as follows
\begin{align*}
\partial \Delta_{1}/\partial \pi_{ij} & =  -\sum_{k\neq j,~J} \frac{2c_{ik}}{\pi_{i+}} - \sum_{k\neq j,~J} \frac{c^{2}_{ik}}{\pi^{2}_{i+}\pi_{+k}}, \\
\partial \Delta_{2}/\partial \pi_{ij}  & = -\sum_{m\neq i,~I} \frac{2c_{mj}}{\pi_{+j}} - \sum_{m\neq i,~I} \frac{c^{2}_{mj}}{\pi_{m+}\pi^{2}_{+j}}, \\
\partial \Delta_{3}/\partial \pi_{ij}  & = \frac{2c_{ij}}{\pi_{i+}\pi_{+j}} - \frac{2c_{ij}}{\pi_{i+}} - \frac{2c_{ij}}{\pi_{+j}} - \frac{c_{ij}^2}{\pi^{2}_{i+}\pi_{+j}} - \frac{c_{ij}^2}{\pi_{i+}\pi^{2}_{+j}} , \\
\partial \Delta_{4}/\partial \pi_{ij}  & = \frac{2c_{iJ}}{\pi_{+J}} - \frac{2c_{iJ}}{\pi_{i+}} + \frac{c_{iJ}^2}{\pi_{i+}\pi^{2}_{+J}} - \frac{c_{iJ}^2}{\pi^{2}_{i+}\pi_{+J}}, \\
\partial \Delta_{5}/\partial \pi_{ij}  & = \frac{2c_{Ij}}{\pi_{I+}} - \frac{2c_{Ij}}{\pi_{+j}} + \frac{c_{Ij}^2}{\pi^{2}_{I+}\pi_{+j}} - \frac{c_{Ij}^2}{\pi^{2}_{+j}\pi_{I+}}, \\
\partial \Delta_{6}/\partial \pi_{ij}  & =  \sum_{k\neq j,~J} \frac{2c_{Ik}}{\pi_{I+}} + \sum_{k\neq j,~J} \frac{c^{2}_{Ik}}{\pi^{2}_{I+}\pi_{+k}}, \\
\partial \Delta_{7}/\partial \pi_{ij}  & = \sum_{m\neq i,~I} \frac{2c_{mJ}}{\pi_{+J}} + \sum_{m\neq i,~I} \frac{c^{2}_{mJ}}{\pi_{m+}\pi^{2}_{+J}}, \\
\partial \Delta_{8}/\partial \pi_{ij} & = -\frac{2c_{IJ}}{\pi_{I+}\pi_{+J}} + \frac{2c_{IJ}}{\pi_{I+}} + \frac{2c_{IJ}}{\pi_{+J}} + \frac{c_{IJ}^2}{\pi^{2}_{I+}\pi_{+J}} + \frac{c_{IJ}^2}{\pi_{I+}\pi^{2}_{+J}}.
\end{align*}
Summarizing the results above, we have 
\begin{equation*}
\Delta '_{ij} = \sum_{k=1}^{J}\frac{2c^{2}_{Ik}}{\pi^{2}_{I+}\pi_{+k}} + \sum_{m=1}^{I}\frac{2c^{2}_{mJ}}{\pi_{m+}\pi^{2}_{+J}} - \sum_{k=1}^{J}\frac{2c^{2}_{ik}}{\pi^{2}_{i+}\pi_{+k}} -  \sum_{m=1}^{I}\frac{2c^{2}_{mj}}{\pi_{m+}\pi^{2}_{+j}} + \frac{2c_{ij}}{\pi_{i+}\pi_{+j}} - \frac{2c_{IJ}}{\pi_{I+}\pi_{+J}},  \text{  for } i \neq I, j \neq J. 
\end{equation*}

For $i = I$, $1\leq j\leq J$,  $\Delta$ can be divided into the following four parts that depend on $\pi_{ij}$
\begin{align*}
\Delta_{1} & =  \sum_{m\neq I} \frac{(\pi_{mJ} - \pi_{m+}\pi_{+J})^2}{\pi_{m+}\pi_{+J}}, \\
\Delta_{2} & = \frac{(\pi_{Ij} - \pi_{I+}\pi_{+j})^2}{\pi_{I+}\pi_{+j}}, \\ 
\Delta_{3} & =  \sum_{m\neq I} \frac{(\pi_{mj} - \pi_{m+}\pi_{+j})^2}{\pi_{m+}\pi_{+j}}, \\
\Delta_{4} & = \frac{(\pi_{IJ} - \pi_{I+}\pi_{+J})^2}{\pi_{I+}\pi_{+J}}.
\end{align*}
The partial derivative of each part with respect to $\pi_{ij}$ is given as follows
\begin{align*}
\partial \Delta_{1}/\partial \pi_{ij} & =  \sum_{m\neq I} \frac{2c_{mJ}}{\pi_{+J}} +  \sum_{m\neq I} \frac{c^{2}_{mJ}}{\pi_{m+}\pi^{2}_{+J}} , \\
\partial \Delta_{2}/\partial \pi_{ij} & = \frac{2c_{Ij}}{\pi_{I+}\pi_{+j}} - \frac{2c_{Ij}}{\pi_{+j}} - \frac{c^{2}_{Ij}}{\pi_{I+}\pi^{2}_{+j}}, \\ 
\partial \Delta_{3}/\partial \pi_{ij} & =  - \sum_{m\neq I} \frac{2c_{mj}}{\pi_{+j}} - \sum_{m\neq I} \frac{c^{2}_{mj}}{\pi_{m+}\pi^{2}_{+j}} , \\
\partial \Delta_{4}/\partial \pi_{ij} & =  - \frac{2c_{IJ}}{\pi_{I+}\pi_{+J}} + \frac{2c_{IJ}}{\pi_{+J}} + \frac{c^{2}_{IJ}}{\pi_{I+}\pi^{2}_{+J}}.
\end{align*}
Summarizing the results above, we have 
\begin{equation*}
\Delta '_{ij} = \sum_{m=1}^{I}\frac{2c^{2}_{mJ}}{\pi_{m+}\pi^{2}_{+J}} -  \sum_{m=1}^{I}\frac{2c^{2}_{mj}}{\pi_{m+}\pi^{2}_{+j}} + \frac{2c_{Ij}}{\pi_{I+}\pi_{+j}} - \frac{2c_{IJ}}{\pi_{I+}\pi_{+J}},   \text{  for } i = I, j \neq J. 
\end{equation*}
Similarly, we have 
\begin{equation*}
\Delta '_{ij} = \sum_{k=1}^{J}\frac{2c^{2}_{Ik}}{\pi^{2}_{I+}\pi_{+k}} -  \sum_{k=1}^{J}\frac{2c^{2}_{ik}}{\pi^{2}_{i+}\pi_{+k}} + \frac{2c_{iJ}}{\pi_{i+}\pi_{+J}} - \frac{2c_{IJ}}{\pi_{I+}\pi_{+J}}, \text{  for } i \neq I, j = J. 
\end{equation*}
Under fixed alternatives, $c_{ij}$'s are not all zeros, therefore $\mathrm{vec}^{*}(D')$ is not a zero vector. By multivariate delta method, we have
\begin{equation*}
\sqrt{n}(\widehat{\Delta}_{n} - \Delta)  \xrightarrow{d} N\left\{ 0, ~ [ \mathrm{vec}^{*}(\Delta') ]^{\intercal} \Sigma^{*} \mathrm{vec}^{*}(\Delta') \right\}, 
\end{equation*}
where the asymptotic variance $[ \mathrm{vec}^{*}(\Delta') ]^{\intercal} \Sigma^{*} \mathrm{vec}^{*}(\Delta')$ is strictly positive due to the positive definiteness of $\Sigma^{*}$. 
\end{proof}

\section*{A.3. Proof of Theorem \ref{thm2}}
Similar to the proof for Theorem \ref{thm1}, we first derive $D'_{ij}$. For $i\neq I$, $j\neq J$, the squared distance covariance functional $D$ can be divided into the following parts that depend on $\pi_{ij}$
\begin{align*}
D_{1} & =  \sum_{k\neq j,~J} (\pi_{ik} - \pi_{i+}\pi_{+k})^2 , \\
D_{2} & = \sum_{m\neq i,~I} (\pi_{mj} - \pi_{m+}\pi_{+j})^2, \\
D_{3} & = (\pi_{ij} - \pi_{i+}\pi_{+j})^2, \\
D_{4} & = (\pi_{iJ} - \pi_{i+}\pi_{+J})^2, \\
D_{5} & = (\pi_{Ij} - \pi_{I+}\pi_{+j})^2, \\
D_{6} & =  \sum_{k\neq j,~J} (\pi_{Ik} - \pi_{I+}\pi_{+k})^2, \\
D_{7} & = \sum_{m\neq i,~I} (\pi_{mJ} - \pi_{m+}\pi_{+J})^2, \\
D_{8} & = (\pi_{IJ} - \pi_{I+}\pi_{+J})^2,
\end{align*}
where the partial derivative of each part can be computed as follows
\begin{align*}
\partial D_{1}/\partial \pi_{ij} & =  -2 \sum_{k\neq j,~J} c_{ik}\pi_{+k}, \\
\partial D_{2}/\partial \pi_{ij}  & =  -2 \sum_{m\neq i,~I} c_{mj}\pi_{+j}, \\
\partial D_{3}/\partial \pi_{ij}  & =  2c_{ij}(1-\pi_{i+}-\pi_{+j}), \\
\partial D_{4}/\partial \pi_{ij}  & = -2c_{iJ}(\pi_{+J} - \pi_{i+}) , \\
\partial D_{5}/\partial \pi_{ij}  & =  -2c_{Ij}(\pi_{I+} - \pi_{+j}), \\
\partial D_{6}/\partial \pi_{ij}  & =  \sum_{k\neq j,~J} c_{Ik}\pi_{+k}, \\
\partial D_{7}/\partial \pi_{ij}  & =  \sum_{m\neq i,~I}c_{mJ}\pi_{m+}, \\
\partial D_{8}/\partial \pi_{ij} & =  2c_{IJ}(\pi_{I+} + \pi_{+J}).
\end{align*}
Summarizing the results above, we have
\begin{equation*}
D_{ij}' = 2\sum_{m=1}^{I}\pi_{m+}c_{mJ} + 2\sum_{k=1}^{I}\pi_{+k}c_{Ik} - 2\sum_{m=1}^{I}\pi_{m+}c_{mj} - 2\sum_{k=1}^{I}\pi_{+k}c_{ik} + 2c_{ij} - 2c_{IJ}, \text{  for } i \neq I, j \neq J. 
\end{equation*}
For $i = I$, $1\leq j\leq J$,  $D$ can be divided into the following parts that depends on $\pi_{ij}$
\begin{align*}
D_{1} & =  \sum_{m\neq I} (\pi_{mJ} - \pi_{m+}\pi_{+J})^2, \\
D_{2} & = (\pi_{Ij} - \pi_{I+}\pi_{+j})^2, \\ 
D_{3} & =  \sum_{m\neq I} (\pi_{mj} - \pi_{m+}\pi_{+j})^2, \\
D_{4} & =(\pi_{IJ} - \pi_{I+}\pi_{+J})^2,
\end{align*}
where the partial derivatives are 
\begin{align*}
\partial D_{1}/\partial \pi_{ij} & = 2 \sum_{m\neq I} c_{mJ}\pi_{m+}, \\
\partial D_{2}/\partial \pi_{ij}  & =  2c_{Ij}(1-\pi_{I+}), \\
\partial D_{3}/\partial \pi_{ij}  & =  -2 \sum_{m\neq I} c_{mj}\pi_{m+}, \\
\partial D_{4}/\partial \pi_{ij}  & = -2c_{IJ}(1 - \pi_{I+}).
\end{align*}
Summarizing the results above, we have
\begin{equation*}
D_{ij}' = 2\sum_{m=1}^{I}\pi_{m+}c_{mJ} - 2\sum_{m=1}^{I}\pi_{m+}c_{mj} + 2c_{Ij} - 2c_{IJ}, \text{  for } i = I, j \neq J.
\end{equation*}
Similarly, we have
\begin{equation*}
D_{ij}' = 2\sum_{k=1}^{I}\pi_{+k}c_{Ik} - 2\sum_{k=1}^{I}\pi_{+k}c_{ik} + 2c_{iJ} - 2c_{IJ}, \text{  for } i \neq I, j = J. 
\end{equation*}
By delta method, we have 
\begin{equation*}
\sqrt{n}(\widehat{D}_{n} - D) \xrightarrow{d} N\left\{ 0, ~ [ \mathrm{vec}^{*}(D') ]^{\intercal} \Sigma^{*} \mathrm{vec}^{*}(D') \right\}, 
\end{equation*}
where the asymptotic variance $[ \mathrm{vec}^{*}(D') ]^{\intercal} \Sigma^{*} \mathrm{vec}^{*}(D')$ is strictly positive due to the positive definiteness of $\Sigma^{*}$. Finally, by Lemma \ref{lem1}, we have 
\begin{equation*}
\sqrt{n}(\widehat{D}_{n} - \widetilde{D}_{n}) \xrightarrow{a.s.} 0.
\end{equation*}
By Slutsky's theorem, we have
\begin{equation*}
\sqrt{n}(\widetilde{D}_{n} - D) \xrightarrow{d} N\left\{ 0, ~ [ \mathrm{vec}^{*}(D') ]^{\intercal} \Sigma^{*} \mathrm{vec}^{*}(D') \right\}, 
\end{equation*}
which completes the proof.

\section*{A.4. Proof of Theorem \ref{thm3}}
Under finite samples and fixed alternatives, we consider the second-order Taylor's expansion
\begin{align*}
\sqrt{n}(\Delta_{n} - \Delta) & \approx \sqrt{n}[\mathrm{vec}^{*}(\widehat{\pi}) - \mathrm{vec}^{*}(\pi)]^{\intercal} \mathrm{vec}^{*}(\Delta ') + \frac{1}{2}\sqrt{n}[\mathrm{vec}^{*}(\widehat{\pi}) - \mathrm{vec}^{*}(\pi)]^{\intercal} H^{*}(\Delta)[\mathrm{vec}^{*}(\widehat{\pi}) - \mathrm{vec}^{*}(\pi)]  \\
& = \sqrt{n}[\mathrm{vec}^{*}(\widehat{\pi}) - \mathrm{vec}^{*}(\pi)]^{\intercal} \mathrm{vec}^{*}(\Delta ') + \frac{1}{2\sqrt{n}}\sqrt{n}[\mathrm{vec}^{*}(\widehat{\pi}) - \mathrm{vec}^{*}(\pi)]^{\intercal} H^{*}(\Delta)\sqrt{n}[\mathrm{vec}^{*}(\widehat{\pi}) - \mathrm{vec}^{*}(\pi)] ,
\end{align*}
where $H^{*}(\Delta)$ represents the Hessian matrix of $\Delta$ (excluding $\pi_{IJ}$). By the proof of Theorem \ref{thm1}, the first term converges to a normal distribution. We now derive the distribution of the second term. By Equation \ref{mclt}, we have 
\begin{equation} 
\sqrt{n}\left[ \mathrm{vec}^{*}(\widehat{\pi}) - \mathrm{vec}^{*}(\pi) \right] \xrightarrow{d} N(\mathbf{0},~ \Sigma^{*}),
\end{equation}
where $\Sigma^{*}$ is symmetric, invertible and positive definite. Let $Z = \sqrt{n}(\Sigma^{*})^{-1/2}\left[ \mathrm{vec}^{*}(\widehat{\pi}) - \mathrm{vec}^{*}(\pi) \right]$, and it can be verified that $Z$ is asymptotically multivariate normal with expectation zero and identify covariance matrix. The second term can now be approximated by 
$$\frac{1}{2\sqrt{n}} Z^{\intercal}(\Sigma^{*})^{1/2}H^{*}(\Delta)(\Sigma^{*})^{1/2}Z.$$ 
By spectral theorem, this random quantity has the same distribution as
\begin{equation*}
\frac{1}{2\sqrt{n}} \sum_{g = 1}^{IJ-1} \beta_{g} Z_{g}^{2}, 
\end{equation*}
where $Z_{1},~...,~Z_{IJ-1}$ are $i.i.d.$ standard normal random variables, and $\{\beta_{1},~...,~\beta_{G}\}$ are the eigenvalues of $(\Sigma^{*})^{1/2}H^{*}(\Delta)(\Sigma^{*})^{1/2}$. This completes the proof.

\newpage
\section*{Figures and Tables}
\begin{figure}[!htbp]
\begin{center}
\includegraphics[scale=0.5]{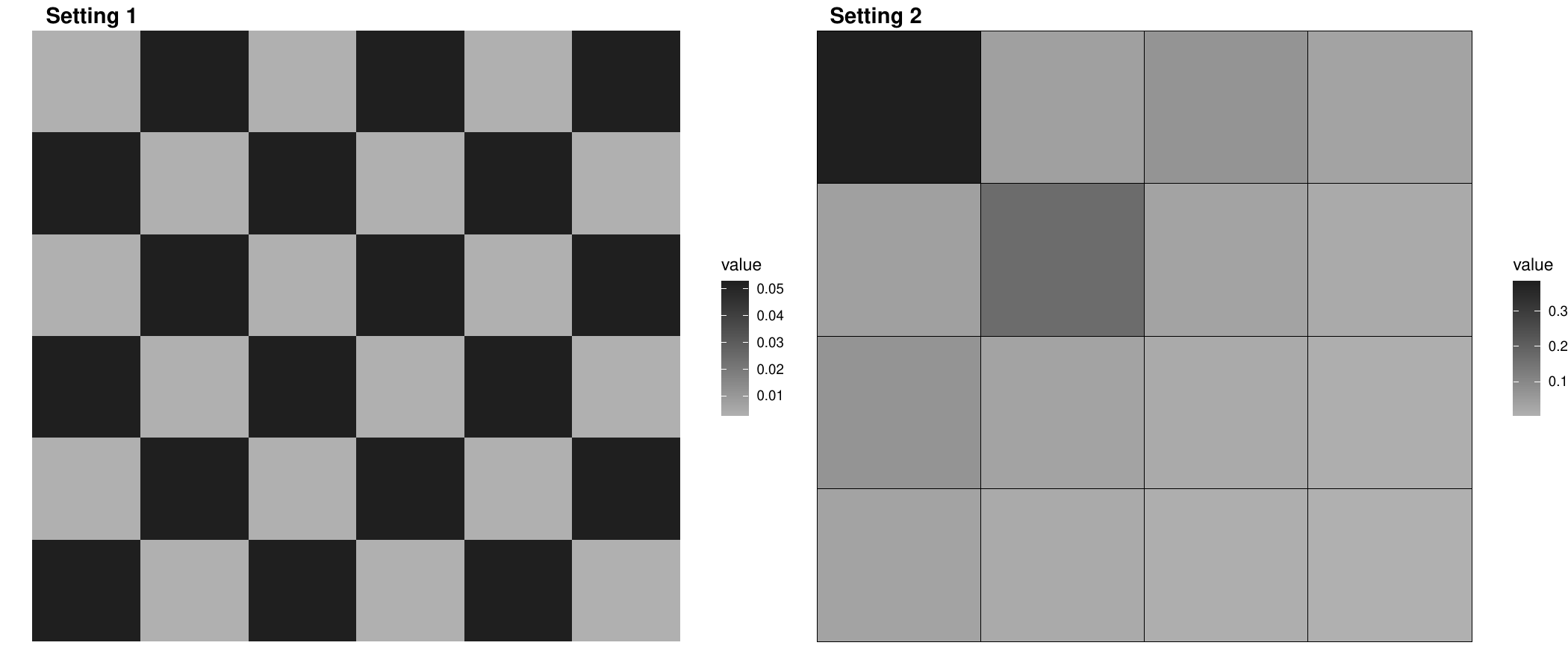}
\end{center}
\caption{Two simulation settings, where $\epsilon = 1/40$ for Setting 1, and $\epsilon = 1/10$ for Setting 2. Darker shades represent higher joint probabilities.
}
\end{figure}

\newpage
\begin{figure}[!htbp]
\begin{center}
\includegraphics[scale=0.5]{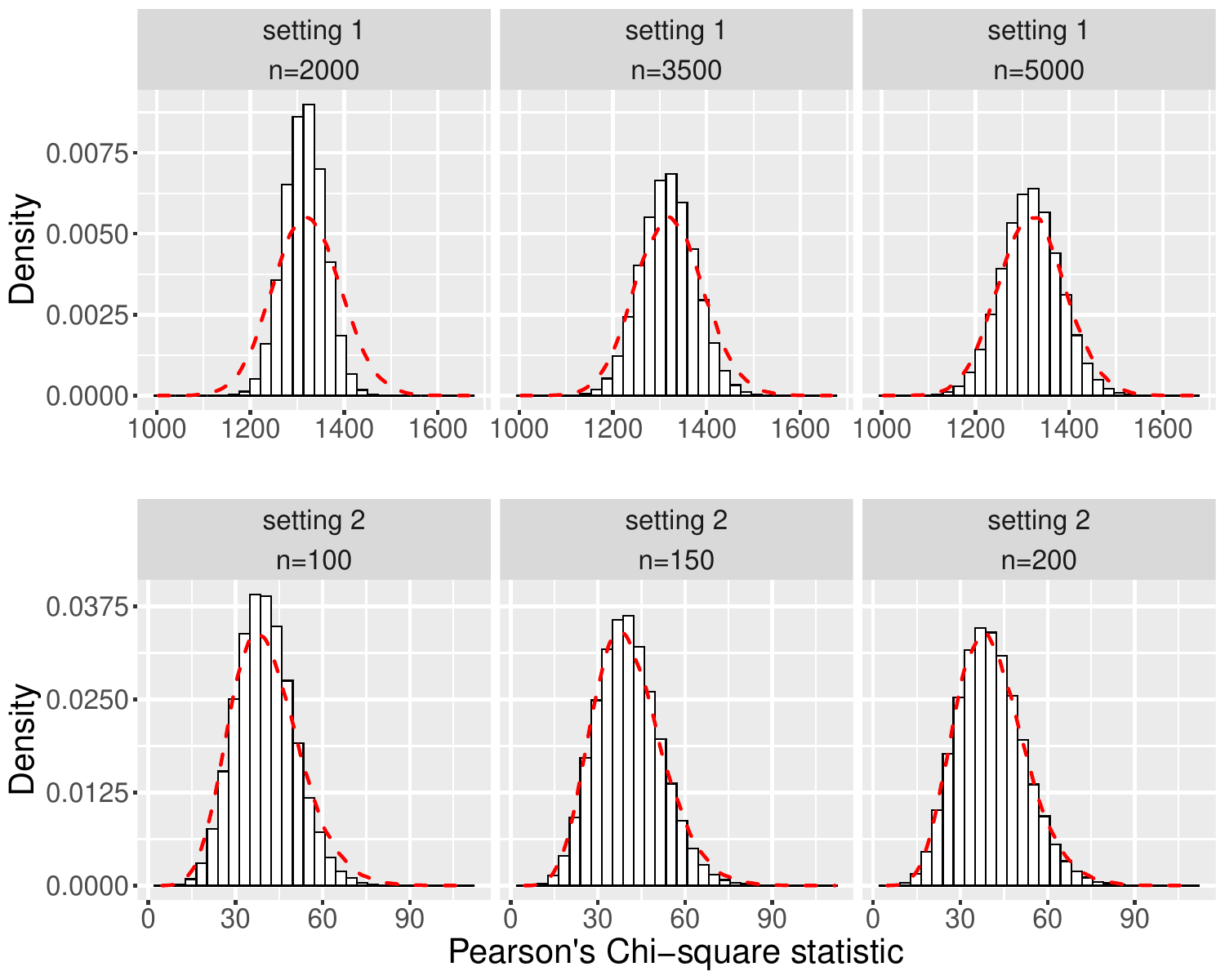}
\end{center}
\caption{Distributions of Pearson's Chi-square statistic ($n\Delta_{n}$) under Pitman's local alternatives ($\epsilon = 1/\sqrt{n}$) with varying sample sizes ($n=2000,~3500,~5000$ for Setting 1 and $n=100,~150,~200$ for Setting 2). The red dashed lines represent the asymptotic distributions derived by Mitra (1958) and Meng \& Chapman (1966).
}
\end{figure}

\newpage
\begin{figure}[!htbp]
\begin{center}
\includegraphics[scale=0.45]{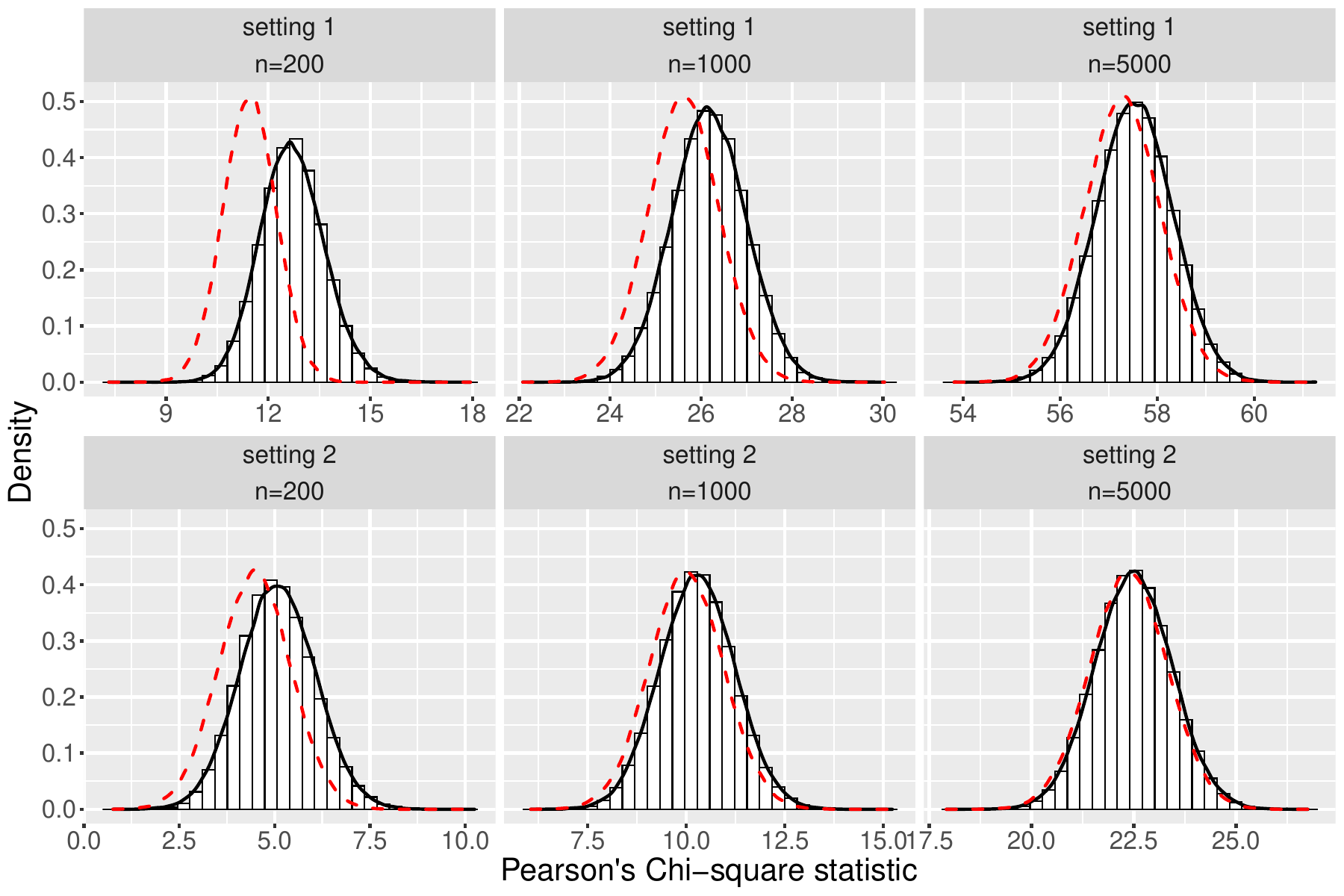}
\end{center}
\caption{Distributions of Pearson's Chi-square statistic ($\sqrt{n}\Delta_{n}$) under fixed alternatives ($\epsilon = 1/40$ for Setting 1 and $\epsilon = 1/10$ for Setting 2) under sample sizes $n=200,~1000,~5000$. The red dashed lines represent the asymptotic distributions derived in Theorems \ref{thm1} and \ref{thm2}, while the black solid lines represent the second-order expansions from Theorems \ref{thm3} and \ref{thm4} (approximated using $100,000$ Monte Carlo samples).
}
\end{figure}

\newpage
\begin{figure}[!htbp]
\begin{center}
\includegraphics[scale=0.45]{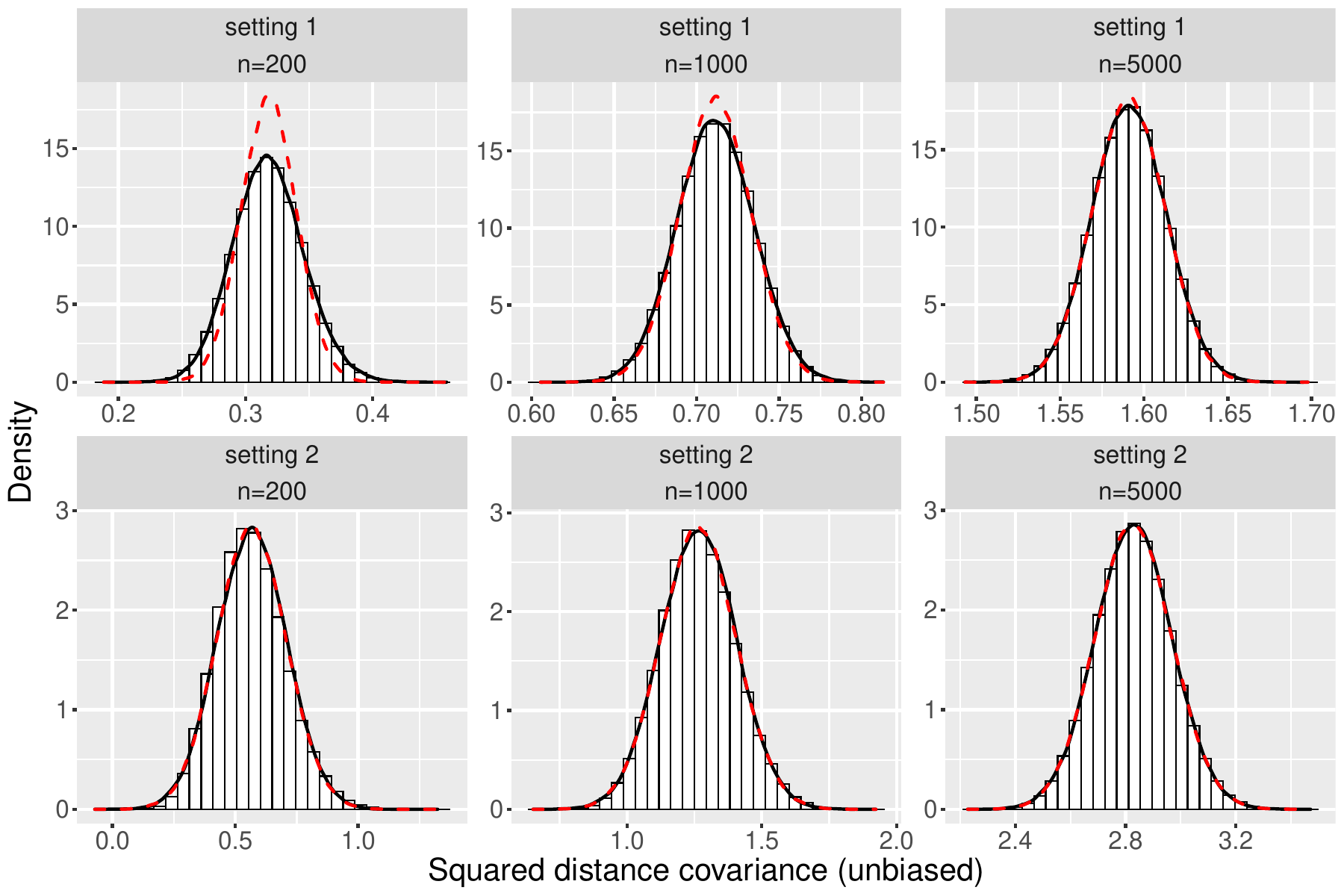}
\end{center}
\caption{Distributions of the unbiased estimate of squared distance covariance ($\sqrt{n}\widetilde{D}_{n}$) under fixed alternatives ($\epsilon = 1/40$ for Setting 1 and $\epsilon = 1/10$ for Setting 2) and sample sizes $n=200,~1000,~5000$. The red dashed lines represent the asymptotic distributions derived in Theorems \ref{thm1} and \ref{thm2}, while the black solid lines represent the second-order expansions from Theorems \ref{thm3} and \ref{thm4} (approximated using $100,000$ Monte Carlo samples).
}
\end{figure}

\newpage
\begin{figure}[!htbp]
\begin{center}
\includegraphics[scale=0.45]{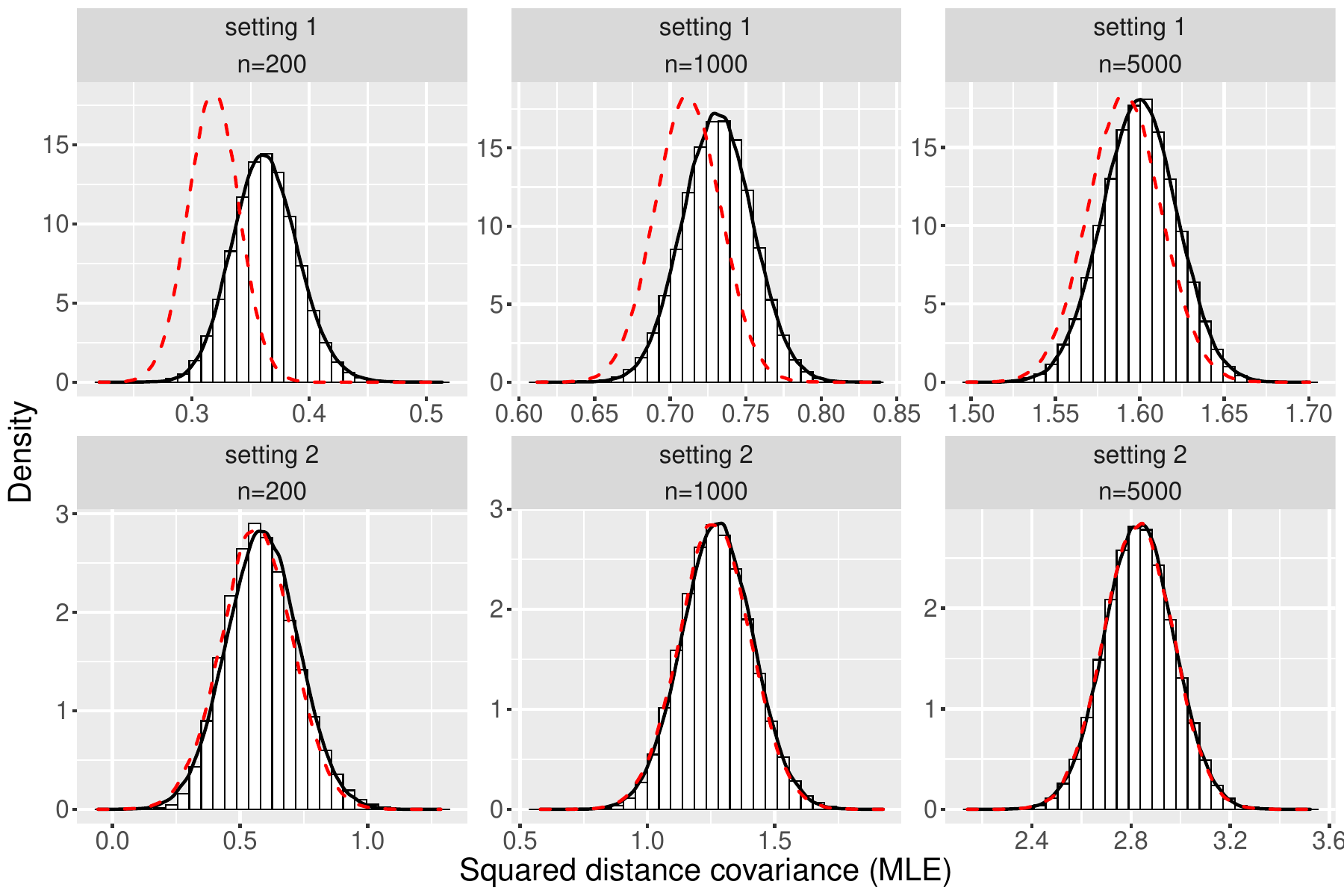}
\end{center}
\caption{Distributions of the maximum likelihood estimate of squared distance covariance ($\sqrt{n}\widehat{D}_{n}$) under fixed alternatives ($\epsilon = 1/40$ for Setting 1 and $\epsilon = 1/10$ for Setting 2) and sample sizes $n=200,~1000,~5000$. The red dashed lines represent the asymptotic distributions derived in Theorems \ref{thm1} and \ref{thm2}, while the black solid lines represent the second-order expansions from Theorems \ref{thm3} and \ref{thm4} (approximated using $100,000$ Monte Carlo samples).
}
\end{figure}

\newpage
\begin{table}[H]
\centering
\title{Table 1: Power comparison for Setting 1}\vspace{2mm}
\rowcolors{1}{}{lightgray}
\begin{tabular}{p{19mm}p{24mm}p{30mm}p{30mm}p{30mm}}
  \hline  \hline 
 $\epsilon$ & n & power$(\widehat{\Delta}_{n})$ & power$(\widehat{D}_{n})$  & power$(\widetilde{D}_{n})$ \vspace{1.5mm} \\
  \hline
1/100 & 100 & 0.449/ 0.445 & 0.482/ 0.464 &  0.485/ 0.468   \vspace{1.5mm}\\ 
        & 150 & 0.690/ 0.693 & 0.713/ 0.701 &  0.714/ 0.704  \vspace{1.5mm}\\ 
        & 200 & 0.846/ 0.862 & 0.861/ 0.864 &  0.860/ 0.867   \vspace{1.5mm} \\ 
        & 250 & 0.931/ 0.946 & 0.936/ 0.947 & 0.936/ 0.948  \vspace{1.5mm} \\ 
 \hline
 1/80 & 100 & 0.698/ 0.714 & 0.733/ 0.722 &  0.729/ 0.721  \vspace{1.5mm}\\ 
        & 150 & 0.910/ 0.927 & 0.920/ 0.931 &  0.919/ 0.934  \vspace{1.5mm}\\ 
        & 200 & 0.977/ 0.987 & 0.980/ 0.987 & 0.980/ 0.989   \vspace{1.5mm} \\ 
        & 250 & 0.994/ 0.998 & 0.996/ 0.998 &  0.996/ 0.998 \vspace{1.5mm} \\ 
  \hline  \hline
\end{tabular}
\caption*{The table above presents the statistical power of three tests under Setting 1. The theoretical power, calculated using Theorems \ref{thm3} and \ref{thm4}, is compared to the simulation-based power (e.g., 0.698/0.714 represents a theoretical power of 0.698 compared to a simulation-based power of 0.714). The values of $\epsilon$ are $1/100$ and $1/80$, and the sample sizes are $100$, $150$, $200$, and $250$.
}
\end{table}

\newpage
\begin{table}[H]
\centering
\title{Table 2: Power comparison for Setting 2}\vspace{2mm}
\rowcolors{1}{}{lightgray}
\begin{tabular}{p{19mm}p{24mm}p{30mm}p{30mm}p{30mm}}
  \hline  \hline 
 $\epsilon$ & n & power$(\widehat{\Delta}_{n})$ & power$(\widehat{D}_{n})$  & power$(\widetilde{D}_{n})$ \vspace{1.5mm} \\
  \hline
1/20 & 100 & 0.482/ 0.450  & 0.748/ 0.726  & 0.743/ 0.721     \vspace{1.5mm}\\ 
        & 150 & 0.681/ 0.656  &  0.862/ 0.878 & 0.864/ 0.878   \vspace{1.5mm}\\ 
        & 200 &  0.810/ 0.807 & 0.923/ 0.953  &  0.922/ 0.952    \vspace{1.5mm} \\ 
        & 250 &  0.887/ 0.897 & 0.955/ 0.983 &  0.955/ 0.983  \vspace{1.5mm} \\ 
 \hline
 1/15 & 100 &  0.763/ 0.754 & 0.905/ 0.926  &  0.907/ 0.929   \vspace{1.5mm}\\ 
        & 150 & 0.910/ 0.923  & 0.964/ 0.988  &   0.963/ 0.989  \vspace{1.5mm}\\ 
        & 200 & 0.967/ 0.981 &  0.986/ 0.997 &  0.985/ 0.998   \vspace{1.5mm} \\ 
        & 250 & 0.989/ 0.996  & 0.995/ 0.999  &  0.996/ 0.999  \vspace{1.5mm} \\ 
  \hline  \hline
\end{tabular}
\caption*{The table above presents the statistical power of three tests under Setting 2. The theoretical power, calculated using Theorems \ref{thm3} and \ref{thm4}, is compared to the simulation-based power (e.g., 0.763/0.754 indicates a theoretical power of 0.763 compared to a simulation-based power of 0.754). The values of $\epsilon$ are $1/20$ and $1/15$, and the sample sizes are $100$, $150$, $200$, and $250$.}
\end{table}

\end{document}